\documentclass{pasj01}
\usepackage{url}

\Received{$\langle$reception date$\rangle$}
\Accepted{$\langle$acception date$\rangle$}
\Published{$\langle$publication date$\rangle$}

\RequirePackage{color}

\begin{document}

\title{Structure of the Milky Way stellar halo out to its outer boundary with
blue horizontal-branch stars}

\author{Tetsuya~Fukushima\altaffilmark{1}, Masashi~Chiba\altaffilmark{1}, 
Daisuke~Homma\altaffilmark{1,3}, 
Sakurako~Okamoto\altaffilmark{3,2}, Yutaka~Komiyama\altaffilmark{3,4}, 
Masayuki~Tanaka\altaffilmark{3}, Mikito~Tanaka\altaffilmark{1}, Nobuo~Arimoto\altaffilmark{5},
and Tadafumi~Matsuno\altaffilmark{3,4}
}

\altaffiltext{1}{Astronomical Institute, Tohoku University, Aoba-ku,
Sendai 980-8578, Japan \\E-mail: {\it t.fukushima@astr.tohoku.ac.jp}}
\altaffiltext{2}{Shanghai Astronomical Observatory, 80 Nandan Road, Shanghai 200030, China}
\altaffiltext{3}{National Astronomical Observatory of Japan, 2-21-1 Osawa, Mitaka,
Tokyo 181-8588, Japan}
\altaffiltext{4}{The Graduate University for Advanced Studies, Osawa 2-21-1, Mitaka,
Tokyo 181-8588, Japan}
\altaffiltext{5}{Astronomy Program, Department of Physics and Astronomy,
Seoul National University, 599 Gwanak-ro, Gwanaku-gu, Seoul, 151-742, Korea}


\maketitle

\begin{abstract}
We present the structure of the Milky Way stellar halo beyond Galactocentric distances
of $r = 50$~kpc traced by blue horizontal-branch (BHB) stars, which are extracted from
the survey data in the Hyper Suprime-Cam Subaru Strategic Program (HSC-SSP). 
We select BHB candidates based on $(g,r,i,z)$ photometry, where the $z$-band is on the
Paschen series and the colors that involve the $z$-band are sensitive to surface gravity.
About 450 BHB candidates are identified between $r = 50$~kpc and 300~kpc, most of which
are beyond the reach of previous large surveys including the Sloan Digital Sky Survey. 
We find that the global structure of the stellar halo in this range has substructures, which are
especially remarkable in the GAMA15H and XMM-LSS fields in the HSC-SSP. 
We find that the stellar halo can be fitted to a single power-law density profile
with an index of $\alpha \simeq 3.3$ ($3.5$) with (without) these fields and
its global axial ratio is $q \simeq 2.2$ ($1.3$).
Thus, the stellar halo may be significantly disturbed and be made in a prolate form
by halo substructures, perhaps associated with the Sagittarius stream in its extension
beyond $r \sim 100$~kpc. For a broken power-law model allowing different power-law indices
inside/outside a break radius, we obtain a steep power-law slope of $\alpha \sim 5$ outside a break
radius of $\simeq 100$~kpc ($200$~kpc) for the case with (without) GAMA15H and XMM-LSS.
This radius of $200$~kpc might be as close as a halo boundary if there is any,
although larger BHB sample is required from further HSC-SSP survey to increase
its statistical significance.
\end{abstract}



\section{Introduction}

Structure and evolution of a faint, diffuse stellar halo surrounding a disk galaxy like
our own Milky Way are still enigmatic, although it is one of the basic, ancient galactic
components. A stellar halo is especially important as it preserves fossil records of
galaxy formation through hierarchical merging and past accretion events because of
its long dynamical time, compared to dynamically well-relaxed, bright disk components.
This is the reason why, despite its tiny fraction of stellar masses in a galaxy
and the difficulty to identify it, a stellar halo has been paid special attention
to researchers since the seminal papers by \citet{Eggen1962}, \citet{Searle1978} and subsequent
studies (see reviews, e.g., \cite{Helmi2008,Ivezic2012,Feltzing2013,Bland-Hawthorn2014}).

While the structure of the Milky Way stellar halo is traced by several means, e.g.,
stellar kinematics, the simple method is to count and map out its bright tracers, such as
red giant-branch (RGB) stars, RR Lyrae (RRL) and blue horizontal-branch (BHB) stars,
which can be observable even at the outskirts of the Milky Way halo.
The latter, RRL and BHB stars are especially advantageous in this purpose,
as their absolute magnitudes and thus distances can be calibrated in a straightforward way.
Based on the assembly and analysis of these halo tracers, some basic structure of the stellar
halo has been revealed out to a few tens kpc and sometimes $r \sim 100$~kpc from the Galactic center;
the stellar halo consists of a general smooth component and irregular substructures (e.g., \cite
{Sluis1998,Yanny2000,Chen2001,Sirko2004,Newberg2005,Juric2008,Keller2008,Sesar2011,Deason2011,Xue2011,
Deason2014,Cohen2015,Cohen2017,Vivas2016,Slater2016,Xu2018,Deason2018,Hernitschek2018}).

The smooth halo component is often modeled as a power-law radial profile with
an index $\alpha$ and an axial ratio $q$. Previous works have attempted to obtain
these density parameters and reached a rough agreement of $2 < \alpha < 4$ and
$0.4 < q < 0.8$, namely rapidly falling density profile with an oblate to nearly
round shape \citep{Sluis1998,Yanny2000,Chen2001,Newberg2005,Juric2008,Sesar2011,Deason2011,
Slater2016,Xu2018}. Most recently, \citet{Hernitschek2018} presented the density
profile of RR Lyrae stars selected from the Pan-STARRS1 survey, which probe the most outer
halo of the Galaxy out to Galactocentric distance of $r = 135$~kpc ever done
using these variables, and obtained $\alpha = 4.4$ and $q = 0.9$ over $20 < r < 135$~kpc.
There are also evidence for a non-monotonous halo structure, such that these halo parameters
vary with radius, from a flattened shape in the inner parts to a less-flattened shape with
a steeper slope in the outer parts \citep{Hartwick1987,Deason2014,
Cohen2015,Cohen2017,Hernitschek2018}.
The stellar halo also shows evidence for a wealth of substructures, especially revealed by
the Sloan Digital Sky Survey (SDSS), including the Sagittarius (Sgr) stream,
Virgo overdensity and the Hercules-Aquila Cloud \citep{Ibata1995,Belokurov2006,
Juric2008}. These lines of evidence suggest that the formation of the stellar halo is indeed
through a series of hierarchical merging/accretion events and this process is continuing perhaps
even in the present day.

While most of the previous works investigate the stellar halo out to 
$r$ of a few tens kpc to $\sim100$~kpc, it is still well below a virial radius
of a MW-sized dark matter halo, typically $200 - 300$~kpc. Also, we have not yet
identified any sharp outer edge of the stellar halo if there is any, so this ancient component
may be much extended without any clear boundary, depending on the recent merging/accretion
history over past billion years \citep{Bullock2005,Deason2014}.
It is thus our motivation of this paper to investigate the structure of the stellar halo
in its outskirts of $r > 100$~kpc. Our work is based on
distant BHB stars from the ongoing Subaru Strategic Program (SSP) using
Hyper Suprime-Cam (HSC) \citep[see for the details of HSC-SSP]{Aihara2018a,Aihara2018b}.
HSC is a new prime-focus camera on Subaru with a 1.5~deg diameter field of view
\citep{Miyazaki2018,Komiyama2018,Furusawa2018,Kawanomoto2018},
whereby enabling sciences with wide and deep imaging data, including
the current work of halo mapping out to its outer boundary.

This paper is organized as follows.
In Section 2, we present the data that we utilize here and the method for the selection
of BHB candidates based on multi-photometry data. The spatial distribution of these BHB
candidates and the method for deriving the radial density profile are also described.
Section 3 is devoted to the results and discussion of our maximum likelihood analysis
for the BHB candidates. We derive the parameters, $\alpha$ and $q$, for the
radial profile of the stellar halo at $r = 50 - 300$~kpc.
Finally, our conclusions are drawn in Section 4.

\begin{table}
\tbl{Observed Regions with HSC-SSP}{%
\begin{tabular}{lccccc}
\hline\hline
Region &   RA  &  DEC  &  $l$  &   b   & adopet area \\
       & (deg) & (deg) & (deg) & (deg) &  (deg$^2$)  \\
\hline
XMM-LSS   &   35  & $-5$ & 170 & $-59$ & 60 \\
WIDE12H   &  180  &    0 & 276 &   60  & 28 \\
WIDE01H   &   19  &    0 & 136 & $-62$ &  0 \\
VVDS      &  337  &    0 &  65 & $-46$ & 48 \\
GAMA15H   &  217  &    0 & 347 &   54  & 85 \\
GAMA09H   &  135  &    0 & 228 &   28  & 90 \\
HECTOMAP  &  242  &   43 &  68 &   47  & 20 \\
AEGIS     &  216  &   51 &  95 &   60  &  2 \\
\hline
\end{tabular}}\label{tab: region}
\end{table}

\section{Data and Method}

\subsection{HSC-SSP data}

This work is based on the imaging data of HSC-SSP survey in its Wide layer,
which is aimed at observing $\sim 1,400$ deg$^2$ in five photometric bands
($g$, $r$, $i$, $z$, and $y$) (for details, see \cite{Aihara2018a,Aihara2018b}).
We use data from the internal s16a data release, which covers
six fields along the celestial equator, named XMM-LSS around at (RA, DEC)
$= (35^{\circ}, -5^{\circ})$, 
WIDE12H at ($180^\circ$, $0^\circ$), WIDE01H at ($19^\circ$, $0^\circ$),
VVDS at ($337^\circ$, $0^\circ$), GAMA15H at ($217^\circ$, $0^\circ$), and
GAMA09H at ($135^\circ$, $0^\circ$) and a field around
(RA,DEC)$=(242^{\circ},43^{\circ})$ (HECTOMAP) as well as a calibration field
around (RA,DEC)$=(216^{\circ},51^{\circ})$ (AEGIS) at the Wide depth, 
amounting to $\sim 300$ deg$^2$ in total (Table \ref{tab: region}).
Since WIDE01H has no $i$ and $z$-band data, we don't use this region. 
The target 5$\sigma$ point-source limiting magnitudes are
($g$, $r$, $i$, $z$, $y$) = (26.5, 26.1, 25.9, 25.1, 24.4) mag.
The HSC data are processed with
hscPipe v4.0.1, a branch of the Large Synoptic Survey Telescope pipeline
\citep{Ivezic2008,Juric2015} calibrated against Pan-STARRS1
photometry and astrometry \citep{Schlafly2012,Tonry2012,Magnier2013}.
All the photometry data are corrected for the mean Galactic foreground
extinction, $A_V$ \citep{Schlegel1998}.

In this paper, for the selection of BHBs by the method described below,
we utilize $g$, $r$, $i$ and $z$-band data for point sources selected using
the {\it extendedness} parameter from the pipeline, namely
{\it extendedness}$=0$ for point sources and {\it extendedness}$=1$ for
extended images like galaxies. For more details of the description of
this parameter, see the data release paper by \citet{Aihara2018b}.
However, this star/galaxy classification becomes uncertain for faint sources.
As detailed in \citet{Aihara2018b}, the contamination, defined as
the fraction of galaxies classified by HST/ACS among HSC-classified stars,
is close to zero at $i<23$, but increases to $\sim 50\%$
at $i=24.5$ at the median seeing of the survey ($0.6$ arcsec). In what follows
of Section 2, we adopt point sources
with $i \le 24.5$ and investigate the possible effect of
the contamination by faint galaxies.

\begin{figure*}[t!]
\begin{center}
\includegraphics[width=160mm]{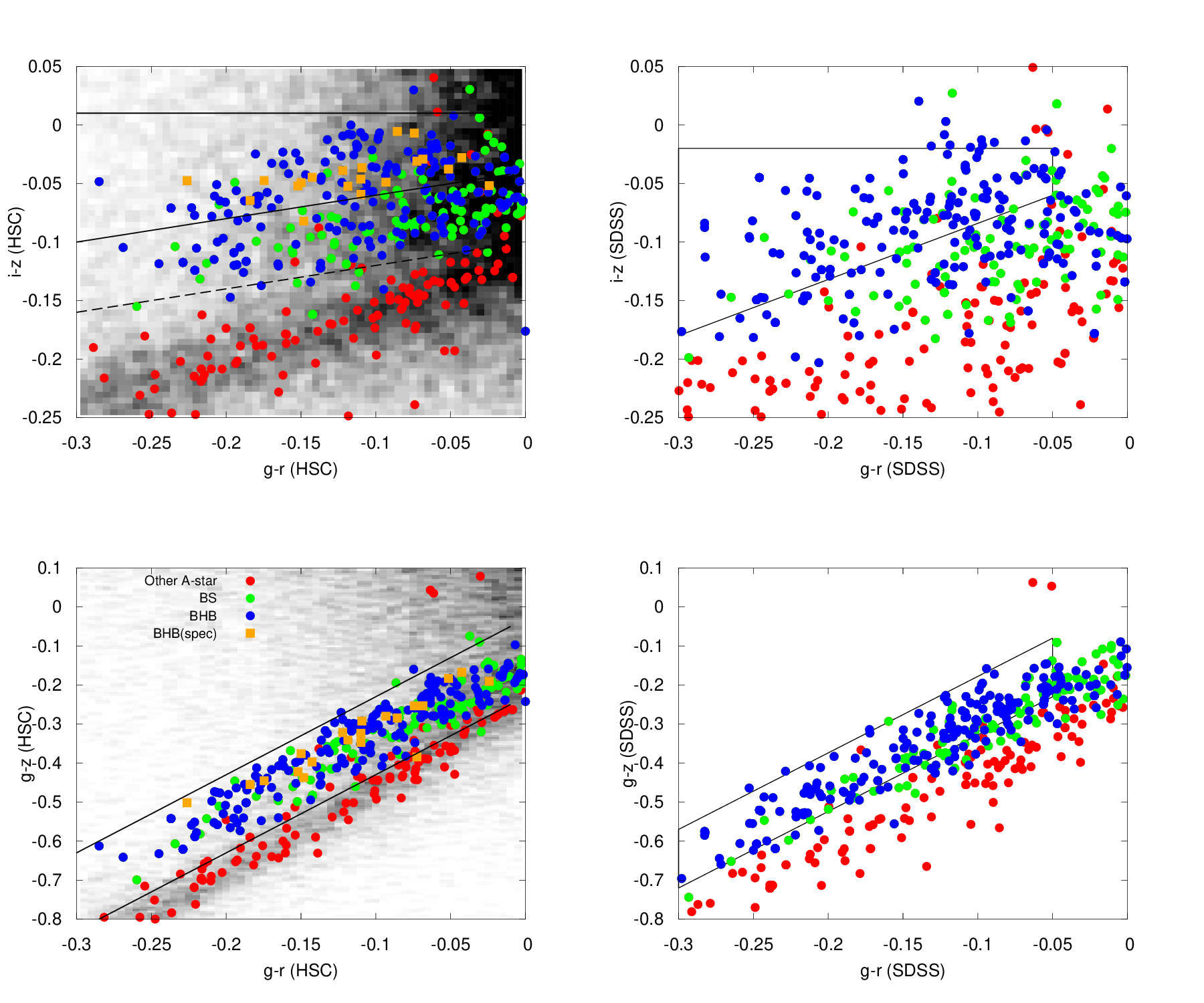}
\end{center}
\hspace*{10mm}
\caption{
Color-color diagrams for the selection of BHB stars in the $g-r$ vs.
$i - z$ space (upper panel) and the $g-r$ vs. $g-z$ space (lower panel).
Left and right panels correspond to the HSC and SDSS filter systems, respectively.
Green and blue points indicate stars classified as BS and BHB candidates,
respectively, whereas red points are other A-colored point sources based on
SDSS's ($u, g, r$)-band selection \citep{Yanny2000}. 
Orange squares show BHB stars selected from spectroscopy by \citet{Xue2011}.
In left panels with the HSC filter system, black dots are all of the point-source data
in HSC-SSP and areas enclosed by solid lines denote the fiducial selection region
for BHB stars against BSs, white dwarfs and quasars adopted in this work.
The dashed lines in the blue side of $i_{\rm SDSS}-z_{\rm SDSS}$ (upper left panel)
is also employed to examine the effects of the contamination by BS stars. 
In the right panel with the SDSS filter system, solid lines show the selection
of BHB stars proposed by \citet{Vickers2012}.
}
\label{fig: color-color}
\end{figure*}

\subsection{Selection of BHB stars}

Candidate BHB stars are often selected using their ultraviolet light as a surface
gravity indicator to distinguish from A-type stars. This is mainly
due to the Balmer jump at 365~nm which is sensitive to surface gravity.
For instance, \citet{Sirko2004} adopt the $u$-band data in the SDSS imaging survey
and set the color cut in the $g-r$ vs. $u-g$ for the selection of BHB stars as
suggested by \citet{Lenz1998}. This method however cannot be applied due to the lack of $u$-band
data in HSC.

\citet{Lenz1998} also suggest the selection in $i-z$ space which is caused by
the Paschen features and is sensitive to surface gravity. \citet{Vickers2012}
develop this selection method using the $z$ band in the $i-z$ vs. $g-r$ diagram
for the removal of A-type stars, white dwarfs and quasars, and also use the $g-z$ vs.
$g-r$ color for the removal of remaining quasars. According to \citet{Vickers2012},
who adopt 10 globular clusters in the SDSS photometry showing pronounced BHBs
for the test, their selection method provides BHBs with $\sim 77$\% pure and $\sim 51$\% complete,
whereas $u$-based color cut selects BHBs with $\sim 74$\% pure and $\sim 57$\% complete.

\citet{Vickers2012} adopted the SDSS filter system to define the 
selection regions of BHBs in both the $i-z$ vs. $g-r$ color and
the $g-z$ vs. $g-r$ color diagrams. Since the $z$-band filter response 
of SDSS is different from that of HSC, we define new selection regions
using the HSC filter system. For this purpose, we select the SDSS photometric data crossmatched
with the HSC data available here, in the restricted color range for A-type stars:
\begin{eqnarray}
-0.3 < g_{\rm SDSS}&-&r_{\rm SDSS} < 0  \\
-0.25< i_{\rm SDSS}&-&z_{\rm SDSS} <0.05 
\label{eq: sdsscolorrange}
\end{eqnarray}
where the latter roughly corresponds to $0< u_{\rm SDSS}-g_{\rm SDSS} <1.5$, which covers
the expected color range for the selection of the BHB stars. 
We also confine ourselves to $g_{\rm SDSS}<20$  to minimize photometric uncertainties.
In Figure \ref{fig: color-color}, we show the $g-r$ vs. $i-z$ and $g-r$ vs. $g-z$ diagrams
for both the HSC and SDSS filter systems, together with BHB and non-BHB candidates taken from
\citet{Yanny2000} based on the $u$-band selection with the SDSS system.
Red points show the sample classified clearly as non-BHB stars, which are located outside
the $u_{\rm SDSS}-g_{\rm SDSS}$ range for BHBs. Both blue and green points are the stars,
which are located within the color cut box with boundaries
$0.8< u_{\rm SDSS}-g_{\rm SDSS} <1.38$ and $-0.3<g_{\rm SDSS}-r_{\rm SDSS}<0.0$, i.e.,
colors occupied by BHBs \citep{Yanny2000}. 
It is well known that these BHB candidates contain blue straggler (BS) stars and
these high-gravity stars are removed based on the further division in the
$u_{\rm SDSS} - g_{\rm SDSS}$ and $g_{\rm SDSS} - r_{\rm SDSS}$ space
\citep{Yanny2000,Deason2011}. Green points denote candidate BSs separated this way,
following the color cut shown in Figure 10 of \citet{Yanny2000}. 
Since this classification method based on the photometric data alone is not so strict,
we also adopt the spectroscopic SEGUE sample of BHB stars compiled by \citet{Xue2011}
and crossmatch this with the current HSC sample. These stars are designated with
orange squares in Figure \ref{fig: color-color}. 
It follows that the BHB candidates selected from the color cut well match those selected
from spectroscopy.

We note from the comparison of the left and right panels in Figure 1 that the HSC system
enables to separate BHBs and non-BHBs more clearly than SDSS. The reason for
this difference is that the HSC $z$-band response is more closely matched with 
the Paschen series than SDSS $z$-band, which are sensitive to the surface gravity.

In this paper, we adopt BHB selection boxes from the HSC filter system as bounded
by solid lines in the left panel of Figure \ref{fig: color-color}. 
These solid lines are defined as
\begin{eqnarray}
-0.3 < g_{\rm HSC}&-&r_{\rm HSC} < 0 \\
0.2(g_{\rm HSC}-r_{\rm HSC}) - 0.04 < i_{\rm HSC}&-&z_{\rm HSC} < 0.01 \\
2(g_{\rm HSC}-r_{\rm HSC}) - 0.23 < g_{\rm HSC}&-&z_{\rm HSC}  < 2(g_{\rm HSC}-r_{\rm HSC}) - 0.03  .
\label{eq: HSCcolorrange}
\end{eqnarray}
We note that this boundary well covers the spectroscopic sample of BHBs.

In addition, we also examine another selection box with a larger area
bounded by a dashed line in the blue side of $i-z$, to investigate the 
effects of the contamination of BS stars (designated by green points) in the later subsection.
In this case, the solid line given as eq.(4) is replaced by the dashed line given as,
\begin{eqnarray}{}
0.2(g_{\rm HSC}-r_{\rm HSC}) - 0.1 < i_{\rm HSC}-z_{\rm HSC} < 0.01 
\label{eq: dashline}
\end{eqnarray}

For the crossmatch, we convert the current HSC filter system to
the SDSS one by the formula given as
\begin{eqnarray}
g_{\rm HSC} &=& g_{\rm SDSS} - a (g_{\rm SDSS} - r_{\rm SDSS}) - b   \label{eq: conversion-g} \\
r_{\rm HSC} &=& r_{\rm SDSS} - c (r_{\rm SDSS} - i_{\rm SDSS}) - d   \\
i_{\rm HSC} &=& i_{\rm SDSS} - e (r_{\rm SDSS} - i_{\rm SDSS}) + f   \\
z_{\rm HSC} &=& z_{\rm SDSS} + g (i_{\rm SDSS} - z_{\rm SDSS}) - h    ,
\label{eq: conversion}
\end{eqnarray}
where $(a,b,c,d,e,f,g,h) = (0.074, 0.011, 0.004, 0.001, 0.106, 0.003, 0.006, 0.006)$ and
the subscript HSC and SDSS denote the HSC and SDSS system, respectively.
These formula, derived by M. Akiyama (private communication, see also
\citet{Homma2016}), have been calibrated from both filter curves and spectral atlas
of stars \citep{Gunn1983}.

\subsection{Contamination of BS stars}

The color cuts given in Figure \ref{fig: color-color} are aimed at clearly separating
and selecting BHB stars, but the color-color space defined for these stars suffers
from finite contamination from BS stars and other populations to some extent. 
We thus need to consider and quantify the effects 
of the contaminants in our selection of BHB stars.
For this purpose, we adopt multi-color ($griz$) HSC photometry of an old stellar system
such as a globular cluster or dwarf spheroid,
from which we select both BHB and BS stars and investigate the efficiency of separating
BHB stars using the color cuts given in Figure \ref{fig: color-color}. In this method, 
we assume that member stars in an old stellar system have similar population properties to
those of field halo stars, which we regard is a reasonable working hypothesis.

The Wide layer in the HSC-SSP covers the area containing a dwarf spheroidal galaxy, 
Sextans, having an extended stellar distribution. We thus adopt this galaxy data for the 
current purpose. So far, yet only the $grz$ imaging data are available in the current HSC-SSP
data set, so to supplement the remaining $i$-band data, we utilize the $gi$-band HSC data of this 
galaxy taken by our group in the Subaru open-use observing program (Chiba et al. S14B-060I).
The cross-matching is made between this and HSC-SSP data using $g$-band photometry for 
Sextans and the candidate member stars of this galaxy spread over its nominal tidal 
radius, $r_t=83.2$~arcmin, are retrieved with the central position of $({\rm RA}, {\rm 
DEC}) = (10:13:02.29, -01:36:53.0)$, position angle of PA$=57.5$~deg and ellipticity of 
$e = 1-b/a=0.29$ \citep{Roderick2016}.

\begin{figure*}[t!]
\begin{center}
\includegraphics[width=80mm]{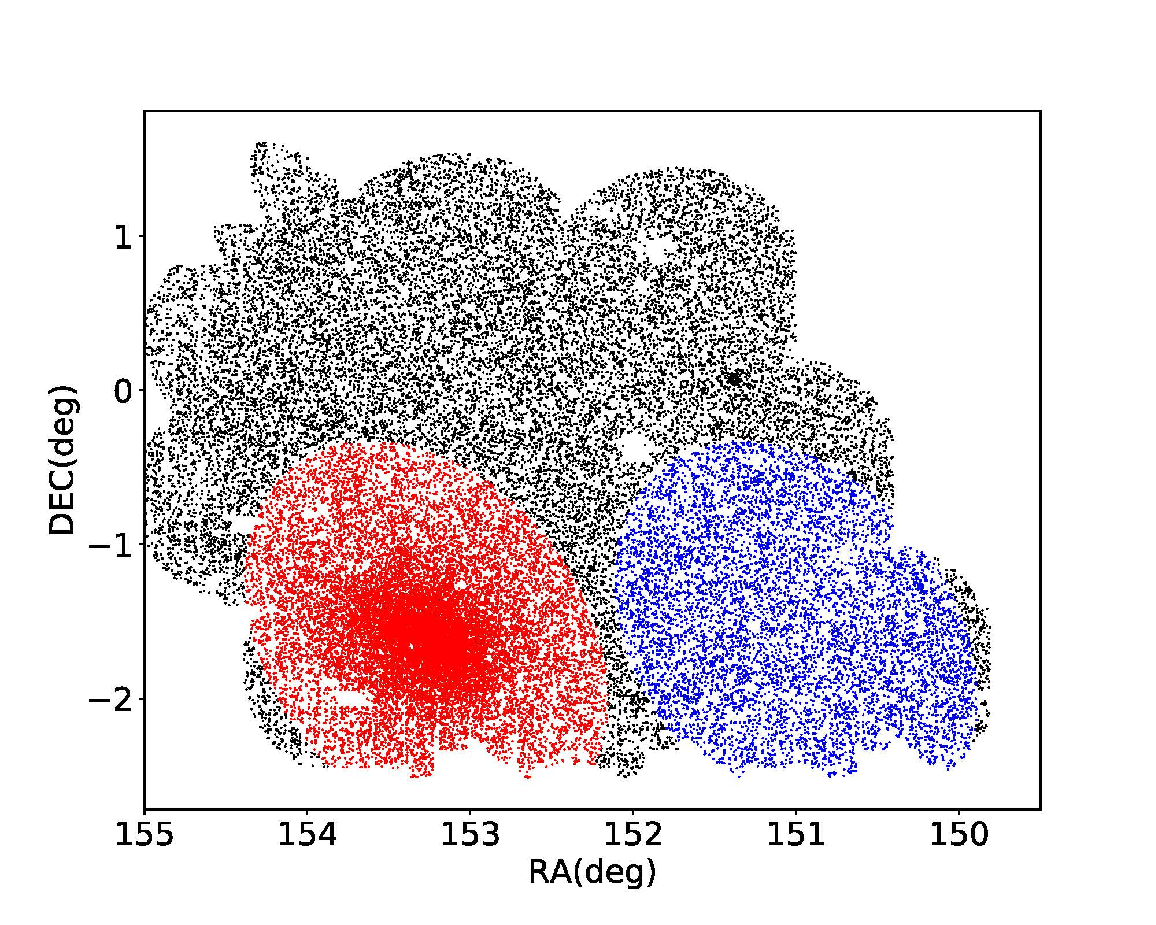}
\includegraphics[width=80mm]{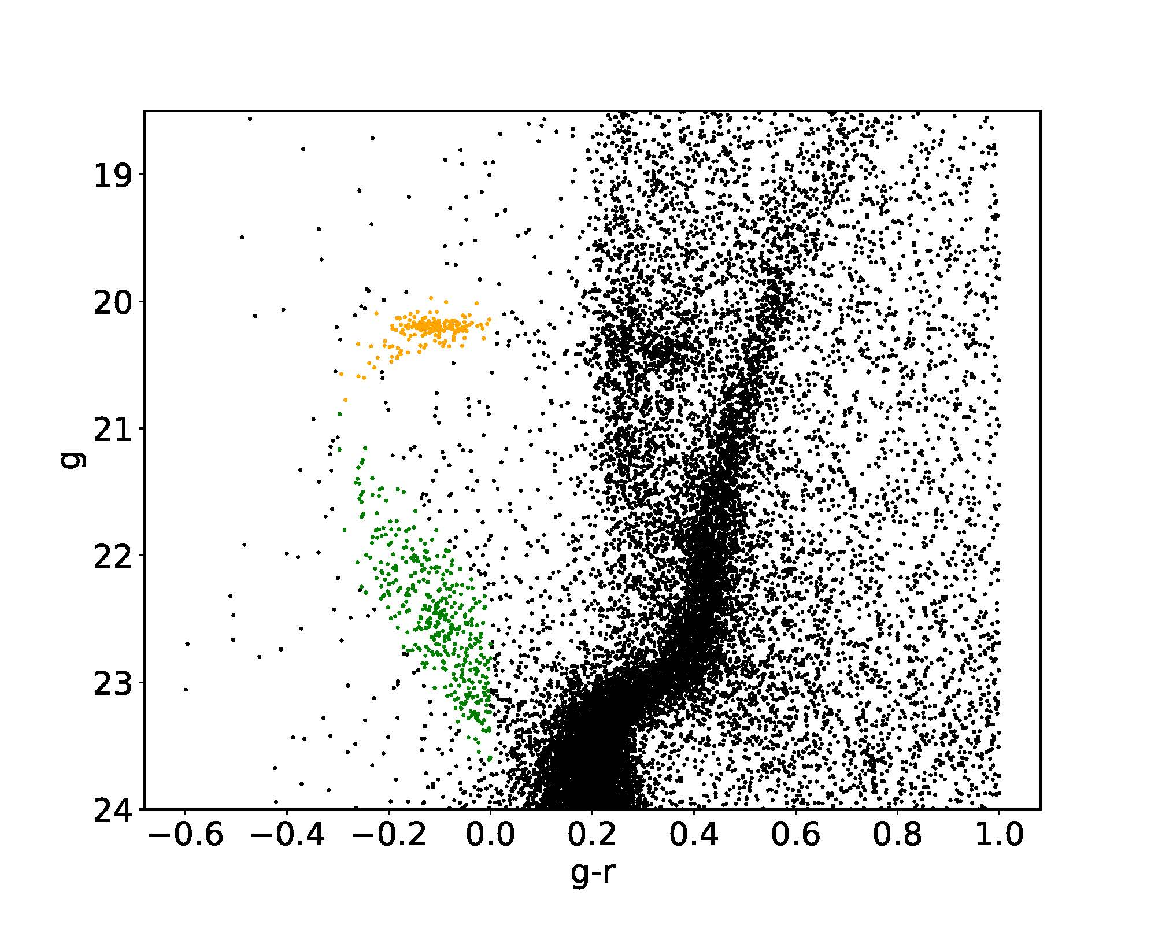}
\end{center}
\hspace*{10mm}
\caption{
Left: spatial distribution of stars in the HSC-SSP data near a dwarf spheroidal galaxy, Sextans.
The candidate member stars in Sextans are designated with red points. Blue points show the
field stars outside Sextans but covering the same area. These field stars are utilized for
the correction in the estimate of selecting BHB stars.
Right: $g$ vs. $g-r$ color diagram of stars in Sextans. Orange and green points
denote the selected BHB and BS stars.
}
\label{fig: Sextans}
\end{figure*}

The left panel of Figure \ref{fig: Sextans} shows the selected regions of Sextans
from HSC-SSP (red points). We also utilize the field stars outside Sextans but
distributed over the same area (blue points) for their correction of the following analysis.
The right panel of Figure \ref{fig: Sextans} shows the $g$ vs. $g-r$ color-magnitude diagram
in Sextans.
We then select candidate BHB and BS stars in 
Sextans at a distance modulus of $m_g - M_g = 19.672$~mag \citep{Roderick2016} as well as for the 
selected field regions, defined as $-0.3 < g-r < 0$ and $19.472 < (m_g - M_g) < 
19.972$ for BHB stars (orange points in the right panel of Figure \ref{fig: Sextans})
and $-0.3 < g-r < 0$ and $19.272 < (m_g - M_g) < 20.572$ for BS stars (green points),
where we use $g$-band absolute magnitudes of BHB and BS stars in equations (\ref{eq: Mg_BHB})
and (\ref{eq: Mg_BS}) as given below.
Next, we set the color cuts defined in Figure \ref{fig: color-color}
for these stars and count the number of each stellar population based on these cuts, as 
summarized in Table \ref{tab: completeness}, where $N_{\rm tot}$ is the total number of
each of the selected BHB and BS stars, whereas $N_{\rm in}$ and $N_{\rm out}$ are the
corresponding number of stars inside/outside the color cuts in Figure \ref{fig: color-color}.
For the selection of BHB stars, we obtain the completeness of
$\sim 67$\% and the purity of $\sim 62$\%. These numbers are compared with those 
for $u$-based color cuts for BHB stars with $\sim 57$\% complete and $\sim 74$\% 
pure \citep{Vickers2012}. It is also worth noting that compared with the use of
the $z$-band photometry by 
SDSS with $\sim 51$\% complete and $\sim 77$\% pure \citep{Vickers2012}, the 
current method using HSC photometry provides a better completeness of selecting BHB 
stars. This is because the HSC $z$-band is more closely matched with the Paschen series
than the SDSS $z$-band.

\begin{table}
\tbl{BHB and BS stars inside/outside Sextans}{
\begin{tabular}{lcccc}
\hline\hline
           & BHB or BS & $N_{\rm tot}$ & $N_{\rm in}$ & $N_{\rm out}$ \\
\hline
Sextans    & BHB       &  178  &  116  &  62  \\
Sextans    & BS        &  411  &   64  & 347  \\
\hline
field      & BHB       &   10  &    3  &   7  \\
field      & BS        &   43  &    2  &  41  \\
\hline
\end{tabular}}
\label{tab: completeness}
\end{table}

\subsection{Distance estimate and spatial distribution of BHBs}

We adopt the formula for $g$-band absolute magnitudes of BHBs, $M_g^{\rm BHB}$,
calibrated by \citet{Deason2011},
\begin{eqnarray}
M_g^{\rm BHB} &=& 0.434 - 0.169(g_{\rm SDSS}-r_{\rm SDSS})  \nonumber \\
      & &+ 2.319(g_{\rm SDSS}-r_{\rm SDSS})^2  + 20.449(g_{\rm SDSS}-r_{\rm SDSS})^3  \nonumber \\
     & &+ 94.517(g_{\rm SDSS}-r_{\rm SDSS})^4  ,
\label{eq: Mg_BHB}
\end{eqnarray}
where both $g$ and $r$ band magnitudes are corrected for interstellar absorption.
To estimate the absolute magnitude of BHBs selected from the HSC data, we also
use eq.(\ref{eq: conversion-g}) - (\ref{eq: conversion}) to translate HSC to SDSS filter system.
We then estimate the heliocentric distances and the three dimensional positions of BHBs
in rectangular coordinates, $(x,y,z)$, for the Milky Way space, where the Sun is
assumed to be at (8.5,0,0)~kpc. To consider the finite effect of contamination from BS stars
as shown below, we adopt their $g$-band absolute magnitudes, $M_g^{\rm BS}$, given by \citet{Deason2011},
\begin{equation}
M_g^{\rm BS} = 3.108 + 5.495 (g_{\rm SDSS}-r_{\rm SDSS}) .
\label{eq: Mg_BS}
\end{equation}

\begin{figure*}
\begin{center}
\includegraphics[width=160mm]{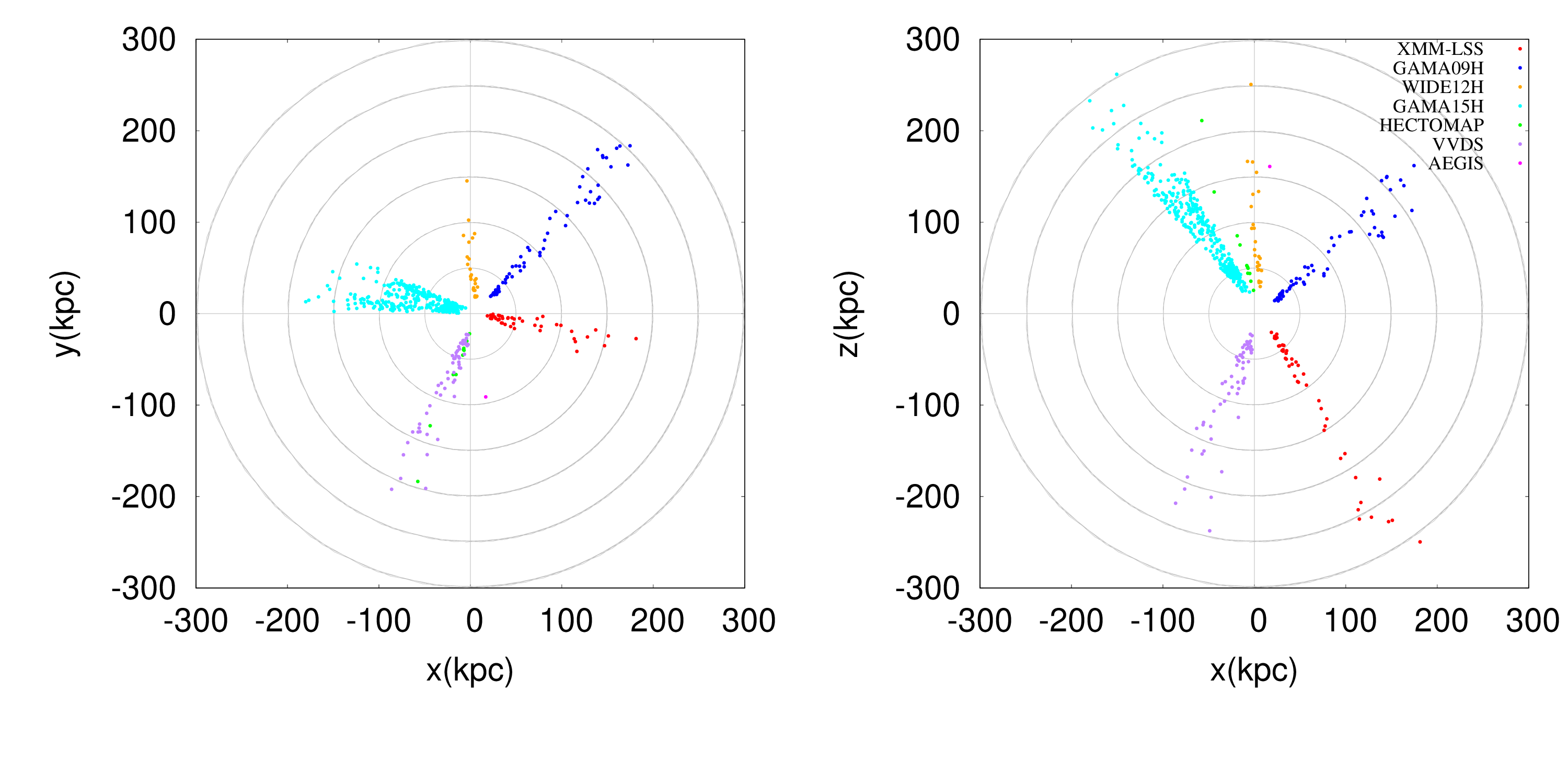}
\end{center}
\hspace*{10mm}
\caption{
The spatial distribution of BHB candidates on the Milky Way $x-y$ plane (left panel)
and $x-z$ plane (right panel) , 
where the Sun is located at $(8.5,0,0)$~kpc and the Galactic plane is defined by the $xy$ plane.
BHBs in each of the survey fields are plotted with different colors.
Here we only use $i \le 23$ mag sample which excludes the contamination of galaxies.
}\label{fig: 3D}
\end{figure*}

Figure \ref{fig: 3D} shows the three dimensional map of BHB candidates in the current sample.
Different colors denote different survey fields. As is clear, the area in each survey
region is yet limited to $\sim 50$~deg$^2$, so the selected BHB stars are distributed
within a pencil cone; AEGIS is confined to the smallest region for its calibration purpose,
so only one BHB is identified in this field.

\begin{figure}[t!]
\begin{center}
\includegraphics[width=80mm]{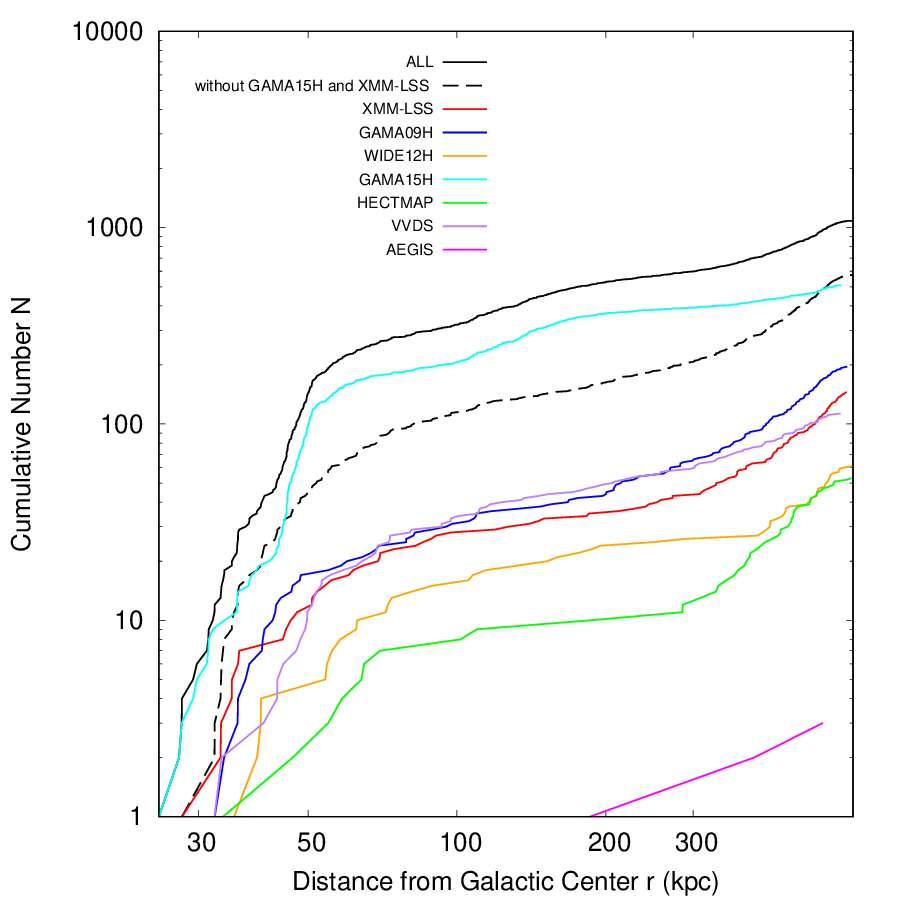}
\end{center}
\hspace*{15mm}
\caption{
Cumulative number distribution of BHB candidates as a function of the radial
distance, $r$, from the center.
We note that an excess of BHB candidates at $r$ beyond $\sim 300$~kpc is
due to the contamination of faint galaxies in the sample. Here we adopt
the faint data with $i \le 24.5$~mag to demonstrate the effect of contamination
from background galaxies.
}
\label{fig: cumu_dist}
\end{figure}

In the GAMA15H field, there exists the so-called Virgo overdensity covering a distance from
$6$ to $44$~kpc and beyond \citep{Juric2008,Vivas2016}, which yields the higher number density
of BHBs than in other fields.
As shown below (cyan line for GAMA15H in Figure \ref{fig: cumu_dist}), in addition to the structure
associated with the Virgo overdensity, we find a secondary structure at $r = 100-200$~kpc,
which would largely affect the determination of the smooth-halo structure.
Also, it is noted that the XMM-LSS field includes a part of the bright stream which exists at
$r = 20-40$~kpc \citep{Koposov2012}.
However, as also shown below (red line in Figure \ref{fig: cumu_dist}), such a substructure
does not clearly appear in the current sample, because our survey region is basically
beyond the corresponding radial range. Since there may exist some unavoidable effects
from this field, we conservatively exclude not only GAMA15H but also XMM-LSS from
the sample when we examine the effects of these known halo substructures on the determination
of the density profile of the halo.

Figure \ref{fig: cumu_dist} shows the cumulative number distribution $N(<r)$ of BHB
candidates as a function of the radial distance from the center, $r$, in each of the
survey fields (colored curve). Black curve shows the distribution by summing up
all fields. Several characteristic features are notable as summarized below.
\begin{itemize}
\item In all fields, there exists an excess of BHB candidates at $r$ beyond $\sim 300$~kpc,
which corresponds to $g$-band magnitude fainter than $\sim 23$~mag or
$i$-band magnitude fainter than $\sim 23.4$~mag at which galaxy contamination
starts to come in \citep{Aihara2018b}. This suggests that the excess feature is due to
the contamination of faint galaxies and that to avoid this contamination effect, we should
confine ourselves to BHB candidates with $i \le 23$~mag or $r \le 300$~kpc.
\item In all fields, there exists a lack of BHB candidates at $r$ below $\sim 30$~kpc,
which corresponds to $g$-band magnitude brighter than $\sim 18$~mag. Note that
such bright objects are often saturated in the HSC-SSP data \citep{Aihara2018b}.
\item GAMA15H shows the highest cumulative number of BHB candidates most probably
due to the presence of halo substructures including the Virgo overdensity and beyond.
\item All fields show similar radial profiles in general.
\end{itemize}

\begin{figure}[t!]
\begin{center}
\includegraphics[width=80mm]{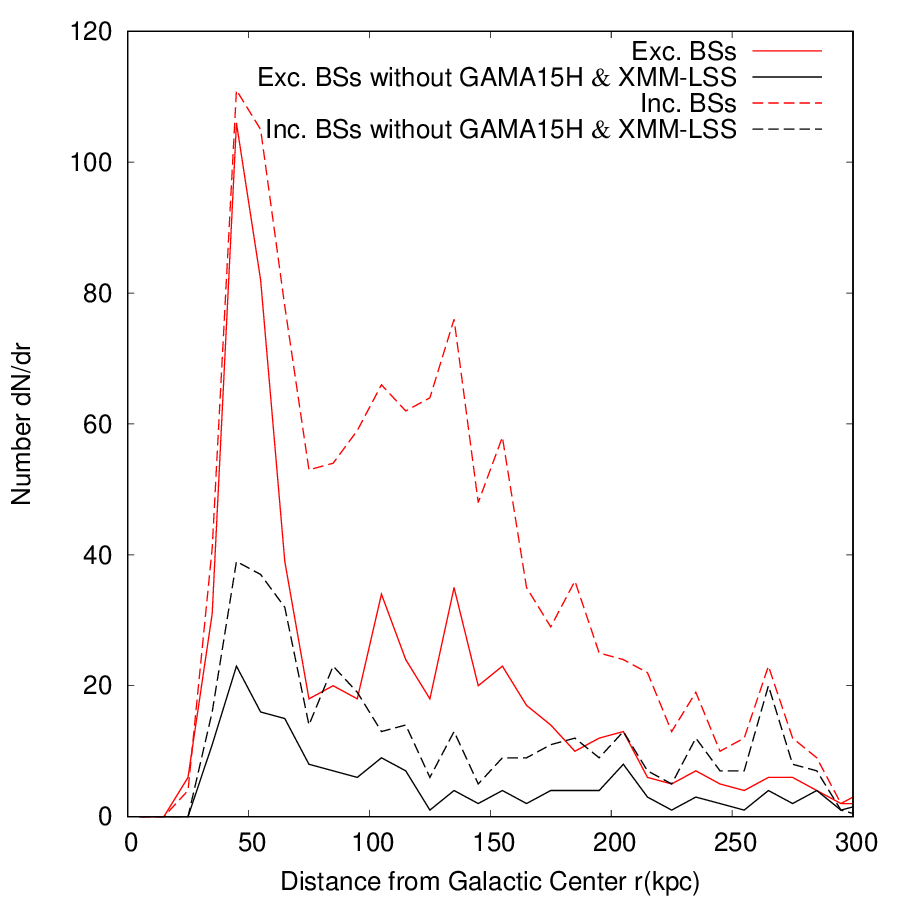}
\end{center}
\hspace*{15mm}
\caption{
Differential number distributions of variously selected stellar populations as a function of $r$.
Red (black) solid lines correspond to our BHB candidates inside the selection box bouded by
solid lines in the left panel of Figure \ref{fig: color-color} with (without) GAMA15H and XMM-LSS.
Red (black) dashed lines correspond to both BHB ad BS candidates by adopting
the selection box bounded by a dashed line in Figure \ref{fig: color-color}
with (without) GAMA15H and XMM-LSS.
}
\label{fig: diff_dist}
\end{figure}

In Figure \ref{fig: diff_dist}, we show the differential distributions of variously selected
stellar populations as a function of $r$, namely the radial density profile $dN(<r)/dr$.
Red (black) solid lines are devoted to our BHB candidates inside the
selection box bounded by solid lines in the left panel of Figure \ref{fig: color-color}
with (without) GAMA15H and XMM-LSS.
There is a peak at $r = 50-60$~kpc, beyond which the sample of BHB candidates
is sufficient enough to enable the derivation of the intrinsic density profile.
This density peak appears to be largely provided by the Virgo overdensity,
since its amplitude is significantly reduced when GAMA15H and XMM-LS
are excluded (black solid line).
Also, we note that the BHB sample with GAMA15H and XMM-LS (red solid line)
shows a secondary structure revealed at $r = 100-200$~kpc, whereas that without
including these fields (black solid line) shows no corresponding feature. This may imply
that the secondary feature at $r = 100-200$~kpc is caused by some finite contamination of
faint BS stars (with about 2 mag fainter luminosities than BHBs) located in
the Virgo overdensity and the bright stream at much inner radii.
To assess this, we consider many of BS candidates in addition to BHB ones by adopting
the selection box bounded by a dashed line in Figure \ref{fig: color-color} and
the results for $dN(r)/dr$ are shown with red (black) dashed lines with (without)
GAMA15H and XMM-LSS in Figure \ref{fig: diff_dist}. It clearly follows that the
secondary feature reported above is much enhanced by including BS candidates,
thereby suggesting that this feature is associated with the substructures including
the Virgo overdensity and bright stream for these faint stars.
We note from the black dashed line that the sample without including
GAMA15H and XMM-LS can exclude the effect of these substructures.

It is also worth remarking that for the case of excluding BS stars without including
GAMA15H and XMM-LSS (black solid line), there is no signature of
a sharp outer edge or rapidly falling density profile beyond $r = 50$~kpc. This is in contrast
to the results of \citet{Deason2014}, who propose, using their BHB sample, a steep
power-law slope at $r$ beyond 50 kpc, i.e., $ \rho \propto r^{-\alpha} $ with $ \alpha \ge 6$,
but in agreement with those of \citet{Cohen2017} using RR Lyrae,
suggesting $\alpha \simeq 4$ for $50 < r < 100$ kpc. This density profile with a power-law
slope of $-3.5$ to $-5$, at least at $r \le 85$~kpc, is also suggested from recent works
by \citet{Slater2016} ($\alpha = 3.5$) and \citet{Xu2018} ($\alpha = 5.0$)
using K giants selected from SDSS and LAMOST, respectively. Most recently, using the
public release of HSC-SSP data over $\sim 100$~deg$^2$ and selecting BHB candidates,
\citet{Deason2018} found a continuation of a $\alpha = 4$ power law from the inner halo
when excluding the Sgr stream even beyond 50~kpc.

Based on these general properties of the sample of BHB candidates, we investigate
their spatial structure in the range of $50 \le r \le 300$ kpc using the sample
with $i \le 23$~mag. We also consider the case with and without including GAMA15H and XMM-LSS
to obtain the effect of substructures in this sample.

\subsection{Maximum likelihood method for getting the radial density profile}

To perform Maximum Likelihood analysis for deriving the most likely radial density 
profile of the BHB stars selected here, while taking account the finite effect of contamination
from BS stars, we adopt and follow the methodology given by 
\citet{Deason2014}. First, based on the experiments for estimating the contaminants 
given above, we define that the membership probabilities of BHB and BS stars, 
$p(griz|{\rm BHB})$ and $p(griz|{\rm BS})$ based on the $griz$ photometry, are 
given as the completeness of including the respective stars in the color cuts.
Second we assume that the ratio between 
the number of BHB and that of BS stars remains constant with magnitude, where the 
fraction of each stellar population relative to the total number of BHB and BS stars is 
given as $f_{\rm BHB}$ and $f_{\rm BS}$, respectively.

Then, for the volume densities of $\rho_{\rm BHB}(m_g-M_g^{\rm BHB}, l, b)$ and 
$\rho_{\rm BS}(m_g-M_g^{\rm BS}, l, b)$ for BHB and BS stars, respectively, we 
define the probability distribution and log-likelihood of
\begin{eqnarray}
P &=& p(griz|{\rm BHB}) \frac{f_{\rm BHB}}{V_{\rm BHB}}
              \rho_{\rm BHB}(m_g-M_g^{\rm BHB}, l, b) D^3_{\rm BHB}  \nonumber \\ 
    & & + p(griz|{\rm BS}) \frac{f_{\rm BS}}{V_{\rm BS}}
              \rho_{\rm BS}(m_g-M_g^{\rm BS}, l, b) D^3_{\rm BS} \\
\log L &=& \sum_{i=1}^{N_{\rm tot}} \log P
\label{eq: likelihood}
\end{eqnarray}
where the subscript $i$ denotes each star in the current sample. $D_{\rm BHB}$ and $D_{\rm BS}$
are distance estimates for BHB and BS stars, respectively, and $V_{\rm BHB}$ and
$V_{\rm BS}$ denote the volumes subtended by the respective stars, which are derived by
integrating over the interval of $18.5 < i < 23$~mag at a color of $(g-r) = -0.05$.
$M_g^{\rm BHB}$ and $M_g^{\rm BS}$ for the absolute magnitudes of BHB and BS stars,
respectively, are given in equations (\ref{eq: Mg_BHB}) and (\ref{eq: Mg_BS}).

In this work, we consider two different models for the radial density profile of BHB 
stars as a halo tracer. The model for a single power-law profile is given in cylindrical 
coordinates $(R,z)$ as
\begin{equation}
\rho(R,z) = \rho_0 R_{\odot}^{\alpha} \left[  R^2 + \frac{z^2}{q^2} 
\right]^{-\alpha/2}  ,
\label{eq: single power density}
\end{equation}
where $\rho_0$ is the density at the position of the Sun $(R,z)=(R_{\odot},0)$ with
$R_{\odot} = 8.5$~kpc, and $\alpha$ and $q$ denote the power-law index and axial 
ratio of the radial density profile, respectively.@Another model is a broken power-law 
profile given as
\begin{equation}
\rho(R,z) = \left\{
\begin{array}{rl}
   \rho_0 R_{\odot}^{\alpha} r_q^{-{\alpha}_{\rm in}}, & 
              \qquad {\rm for} \ r_q \le r_b        \nonumber       \\
   \rho_0 R_{\odot}^{\alpha} r_b^{{\alpha}_{\rm out}-{\alpha}_{\rm in}}
   r_q^{-{\alpha}_{\rm out}}, & \qquad {\rm for} \ r_q > r_b
\label{eq: broken power density}
\end{array} \right.
\end{equation}
where $r_q = \sqrt{R^2 + z^2/q^2}$.

We derive the most likely set of parameters $(\alpha, q)$ for a single power-law model 
and $(\alpha_{\rm in}, \alpha_{\rm out}, r_b, q)$ for a broken power-law model by 
maximizing $L$, $L_{\rm max}$, and estimate their confident intervals from 
$F = -2 \ln L/L_{\rm max}$ provided $F$ has a $\chi^2$ distribution for 2 and 4 degrees of 
freedom for these models, respectively.

\section{Results and Discussion}

We adopt the sample of BHB candidates with $i \le 23$~mag in all survey fields, select those
in the range of $50 \le r \le 300$~kpc, and perform the maximum likelihood analysis as described
in the previous section. The results are summarized in Table \ref{tab: single} and \ref{tab: broken}.

\begin{table*}
\tbl{Maximum Likelihood results for a single power-law model}{
\begin{tabular}{lccc}
\hline\hline
Inclusion of GAMA15H and XMM-LSS & $\alpha$ & $q$ & $\ln L$   \\
\hline
with     & $3.27^{+0.17}_{-0.17}$ & $2.17^{+0.34}_{-0.30}$ & $-110.25$ \\
without  & $3.51^{+0.36}_{-0.40}$ & $1.34^{+0.66}_{-0.38}$ & $   5.24$ \\
\hline
\end{tabular}}
\label{tab: single}
\end{table*}

\begin{table*}
\tbl{Maximum Likelihood results for a broken power-law model}{
\begin{tabular}{lccccc}
\hline\hline
Inclusion of GAMA15H and XMM-LSS & $\alpha_{\rm in}$ & $\alpha_{\rm out}$ & $r_b$ (kpc) & $q$ & $\ln L$   \\
\hline
with     & $3.1^{+0.2}_{-0.5}$ & $4.7^{+0.7}_{-0.9}$ & $105^{+35}_{-25}$ & $2.6^{+1.9}_{-0.4}$ & $-173.2$ \\
without  & $3.2^{+0.9}$        & $5.3^{+0.7}$        & $210$             & $1.5^{+2.1}_{-0.5}$ & $  10.4$ \\
\hline
\end{tabular}}
\label{tab: broken}
\end{table*}

\subsection{The global halo structure over 50 to 300 kpc}

\begin{figure}[t!]
\begin{center}
\includegraphics[width=90mm]{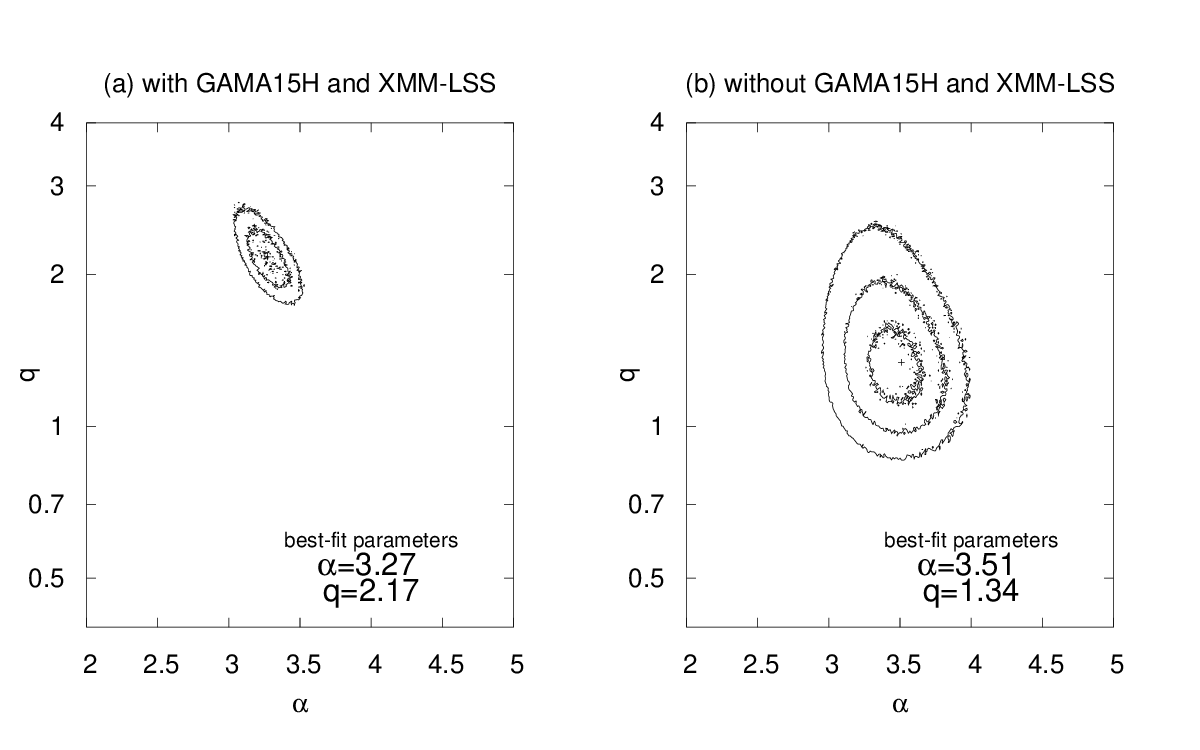}
\end{center}
\hspace*{10mm}
\caption{
Confidence contour plots of the likelihood function $L$ for a single power-law model,
when we consider all the relevant BHB sample
(left panel) and when GAMA15H and XMM-LSS containing notable halo substructures
are excluded (right panel).
Cross shows a best-set of parameters with $L_{\rm max}$ and solid lines show
the 1$\sigma$, 2$\sigma$ and 3$\sigma$ confidence.
}
\label{fig: L_all}
\end{figure}

The left panel in Figure \ref{fig: L_all} shows, for a single power-law model,
confidence contour plots of the likelihood function $L$,
when we consider all the relevant sample with the number $N_{\rm tot}=442$.
There exists clearly a localized maximum at $\alpha \simeq 3.3$ and $q \simeq 2.2$,
suggesting that the stellar halo in this radial range has a largely prolate shape.
On the other hand, the right panel in Figure \ref{fig: L_all} shows the results
when GAMA15H and XMM-LSS having notable substructures are
excluded in the analysis, where $N_{\rm tot}=122$. Although confident intervals are
enlarged due to the small number of the sample, this case reveals the best-fit parameters of
$\alpha \simeq 3.5$ and $q \simeq 1.3$, suggesting that while
the index $\alpha$ remains similar, the shape of the stellar halo becomes rounder.

This result, i.e., the largely prolate shape of the halo when GAMA15H and XMM-LSS
are included, may be due to the presence of notable substructures related to
the Virgo overdensity in GAMA15H. In particular,
these substructures also include a part of the Sgr stream, which is formed
from a tidally disrupting, polar-orbit satellite, Sgr dwarf. 
XMM-LSS also includes a part of the Sgr stream.
Thus, the anisotropic distribution of this tidal stream
may make the stellar halo being prolate in the above fitting process.

To investigate any radial variation of the halo structure in the current sample
of BHB candidates, we also consider a broken power-law model for the halo parameterized
by $(\alpha_{\rm in}, \alpha_{\rm out}, r_b, q)$ (Table \ref{tab: broken}).
It follows that the case with GAMA15H and XMM-LSS yields a change in the density slope
at $r_b \simeq 100$~kpc, beyond which the density profile is steeper
$(\alpha_{\rm out} \simeq 4.7)$ than that in the inner parts
$(\alpha_{\rm out} \simeq 3.1)$. We note that in GAMA15H there exist
halo substructures associated with the Sgr stream extended up to $r \sim 80$~kpc
and this may explain the current result.
On the other hand, without including GAMA15H and XMM-LSS,
we obtain the break radius of $r_b \simeq 200$~kpc and the halo density profile
is made somewhat steeper $(\alpha_{\rm out} \simeq 5.3)$ beyond this radius.
This radius might be as close as a halo boundary if there is any,
which can be formed by the lack of accretion of small galaxies
over the past billion years \citep{Bullock2005,Deason2014},
although this is inferred from yet small number statistics.
For further insight into a halo boundary, we need a much larger sample
with a higher statistical significance, because the BHB sample in outer radii
suffers from misclassification with faint background galaxies. 
Moreover, since the number of the BHB stars by excluding GAMA15H and XMM-LSS
is yet small in the current data set, the associated errors in
$\alpha_{\rm in}$, $\alpha_{\rm out}$ and $r_b$ for this broken power-law model
are large and some of them are actually undetermined in this study (Table \ref{tab: broken}).
Thus, the interpretation of the results for this case still needs a great caution.

It is also worth noting that even in this broken power-law model,
the shape of the stellar halo at $r > 100$~kpc appears largely prolate, especially
when GAMA15H and XMM-LSS are included. This result is compared with
suggested oblate shapes at $r < 50$~kpc derived in previous work, summarized as 
$\alpha \sim 3.2$ and $q \sim 0.5$ for BHBs at $1 < r < 20$~kpc \citep{Sluis1998},
$\alpha \sim 2.6$ and $q \sim 0.65$ for BHBs out $\sim 40$~kpc \citep{Deason2011}, and
$2 < \alpha \le 4$ and $0.4 \le q \le 0.8$ in various other work
\citep{Yanny2000,Chen2001,Newberg2005,Juric2008,Sesar2011}.
This may be understood if there exist some substructures associated with the
Virgo overdensity and a secondary substructure seen beyond 100~kpc.
Indeed, a recent numerical simulation
for investigating the effect of the infalling Sgr dwarf from outside \citep{Dierickx2017}
implies that beyond $r \sim 100$~kpc, the presence of tidal debris associated with
the Sgr stream is predicted in the direction of GAMA15H. This supports the hypothesis that
the larger axial ratio $q$ when GAMA15H is included is due to the effect of the Sgr stream.

\subsection{Comparison with Deason et al. (2018)}

Recently, \citet{Deason2018} presented their analysis of BHB stars using the public
release of the HSC-SSP data over $\sim 100$ deg$^2$. Their method for selecting BHB stars
is basically the same as that adopted in this work using $griz$ multiband photometry
\citep{Vickers2012}, although there are some differences in details in the adopted color cuts
of $i-z$ vs. $g-r$ and $g-z$ vs. $g-r$ as well as the total area of the surveyed regions
used in the analysis, where we make use of the HSC-SSP data over $\sim 300$~deg$^2$ and
thus the total number of identified BHB stars is much larger in our work.

For comparison with their work, we make the number counts of BHB stars following
their Maximum Likelihood method.
Namely, we set, in this work, bins of 0.45 mag in distance modulus over
$18.5 < g - M_g^{\rm BHB} < 23$ and count the number of BHB stars in each bin.
The probability distribution function in each distance modulus bin is defined as
\begin{eqnarray}
P({\bf x}) &=& f_{\rm BHB} p({\bf x}|{\rm BHB}) + f_{\rm BS} p({\bf x}|{\rm BS}) \nonumber \\
       & & + f_{\rm WD} \left[ 0.7p({\bf x}|{\rm WD}_{\rm DA}) + 0.3p({\bf x}|{\rm WD}_{\rm DB}) \right] \nonumber \\
       & & + \frac{f_{\rm QSO}}{{\bf x}_{\rm max} - {\bf x}_{\rm min}}
\label{eq: deason_likelihood1}
\end{eqnarray}
where $p({\bf x}|type)$ is a probability distribution of specified stars in $griz$ space,
which is assumed as Gaussian,
$p({\bf x}|type) = \frac{1}{\sqrt{2\pi\sigma}} \exp(-(x-x_0)^2/2\sigma^2)$ and ${\bf x} = griz$.
The variation of the Gaussian widths, $\sigma$, with magnitude, is given by the sum of the
intrinsic widths and the photometric errors of HSC, 
$\sigma^2 = \sigma_{\rm intrinsic}^2 + \sigma^2(griz)$, where $\sigma_{\rm intrinsic}$ is
kept fixed and taken from Table~1 of \citet{Deason2018}.
Here, the contributions of QSOs and White Dwarfs (WDs) with DA/DB types are given as
their number fractions of $f_{\rm QSO}$ and $f_{\rm WD}$, where the constant contamination from
QSOs are assumed and the ratio between these two types of WDs is set to be $7:3$ \citep{Deason2014}.
The number counts of BHB stars, $N_{\rm BHB}$, are then obtained in each bin
by maximizing the log-likelihood function of
\begin{equation}
\log L = \sum_{i}^{N_{\rm tot}} \log P({\bf x}_i) .
\label{eq: deason_likelihood2}
\end{equation}

\begin{figure}[t!]
\begin{center}
\includegraphics[width=90mm]{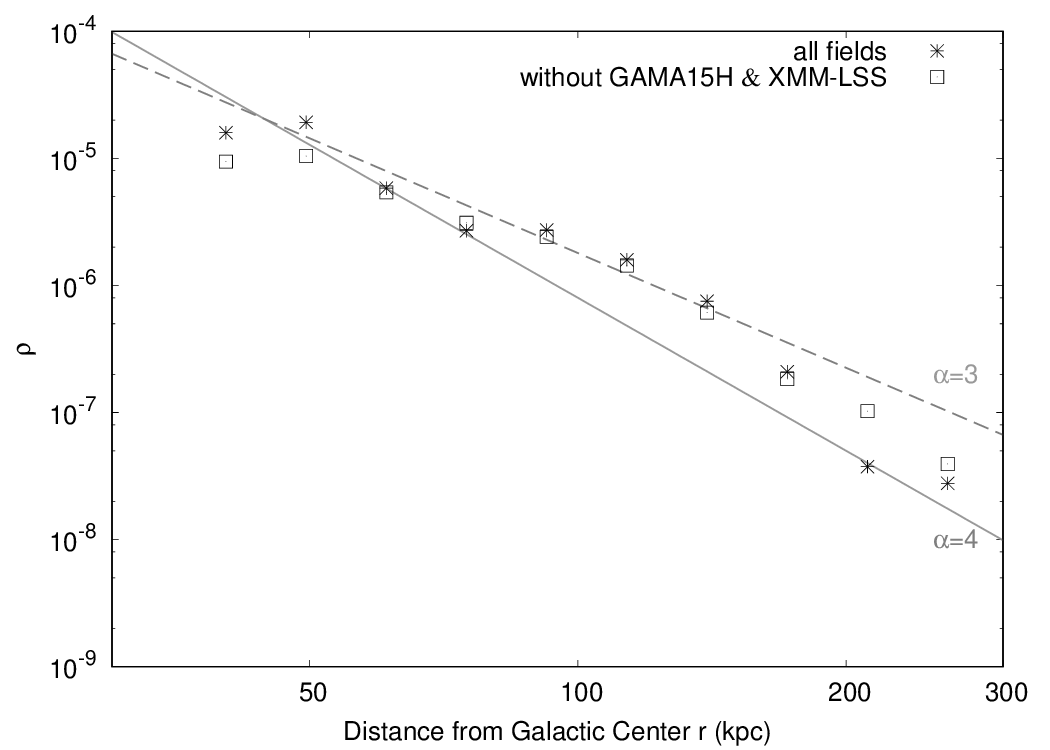}
\end{center}
\hspace*{10mm}
\caption{
Density profiles of BHB stars with (without) GAMA15H and XMM-LSS based on
the Maximum Likelihood method given in equations (\ref{eq: deason_likelihood1}) and
(\ref{eq: deason_likelihood2}) are shown with asterisks (open squares).
Solid and dashed lines show power-law profiles of $\alpha=3$ and $4$, respectively,
for comparison.
}
\label{fig: deason18}
\end{figure}

Figure \ref{fig: deason18} shows the density profile of BHB stars based on this methodology,
where the cases with (without) GAMA15H and XMM-LSS are shown with asterisks (open squares).
It follows that the both cases yield a power-law profile with $\alpha$ being $3$ to $4$;
There is a tendency that beyond a radius at $r \sim 100$ kpc (200 kpc) with (without)
GAMA15H and XMM-LSS, which is basically the same location of a break radius obtained
in the previous subsection, the slope appears steeper than $4$, as also inferred from
the above experiments. For comparison with \citet{Deason2018}, we simply make a $\chi^2$ fitting of
a power-law density profile of $\rho \propto r^{-\alpha}$ to the data over $50 < r < 300$~kpc
and obtain $\alpha = 3.9$ ($3.5$) for the case with (without) GAMA15H and XMM-LSS.
These properties of the current BHB sample are generally in agreement with those in
\citet{Deason2018} reporting $\alpha \sim 4$ from the same analysis and thus we conclude that
both works arrive at basically the same results. We will further examine this broken nature
of the density profile using the future HSC data release.

\subsection{The Sgr stream in GAMA15H and XMM-LSS}

\begin{figure}[t!]
\begin{center}
\includegraphics[width=90mm]{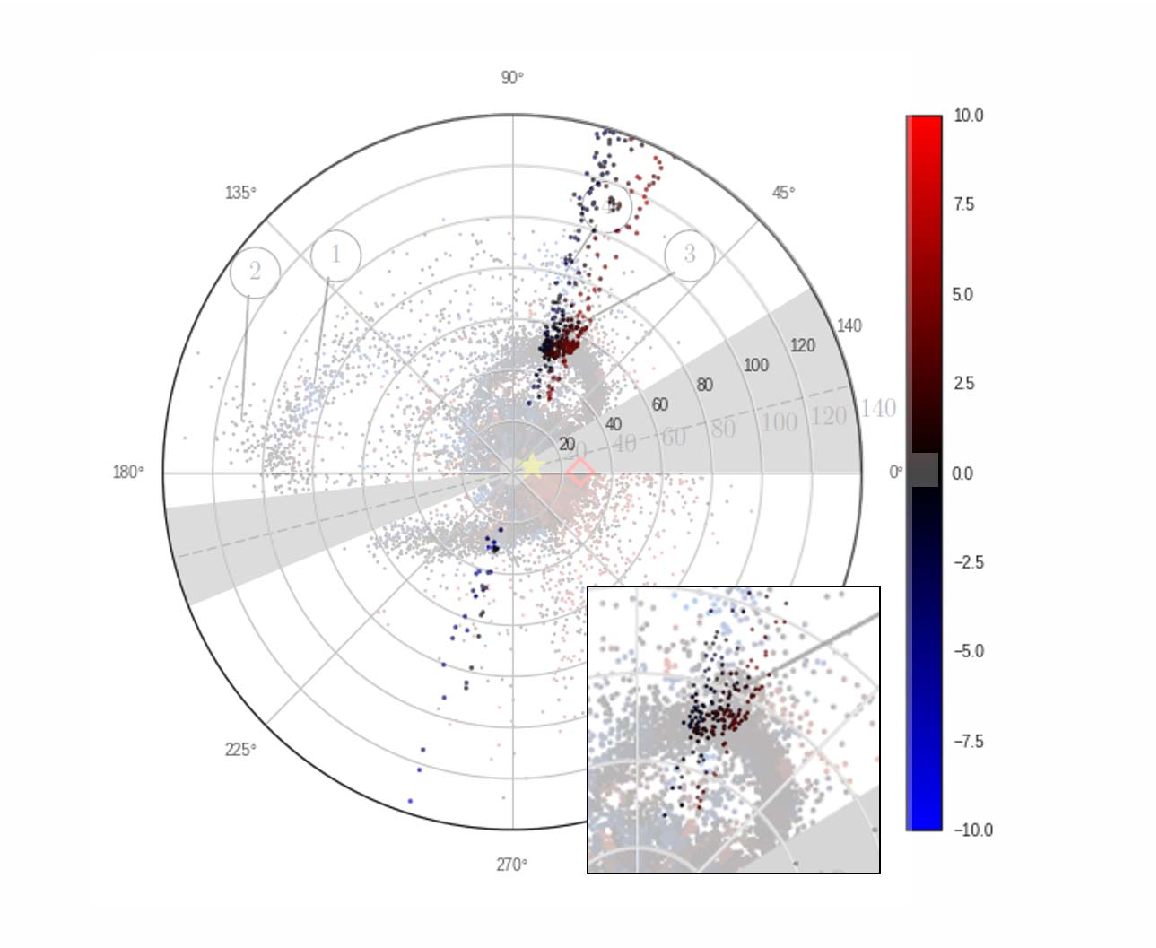}
\end{center}
\caption{
The distribution of BHBs (colored dots) near the Sgr orbital plane
(i.e., $|\tilde{B}_\odot| < 8^\circ$) superposed into the map of PS1 RRab stars
(gray dots) shown in Figure 1 of \citet{Sesar2017}, 
in which substructures associated with the Sgr stream is clearly seen.
The lower right panel shows the enlarged plots in GAMA15H
($60^\circ <\tilde{\Lambda}_\odot < 80^\circ$) near the Sgr stream. 
The color indicates the stars' angular distance from the Sgr orbital plane.
}
\label{fig: stream}
\end{figure}

In previous section we mention that the Sgr stream is present in GAMA15H and XMM-LSS.
Here we investigate the distribution of BHBs relative to that of the Sgr stream in detail.
We adopt the heliocentric Sagittarius coordinates, ($\tilde{\Lambda}_\odot$, $\tilde{B}_\odot$),
as defined by \citet{Belokurov2014}. As shown in Fig. \ref{fig: stream}, BHBs in
GAMA15H and XMM-LSS are distributed near the Sgr orbital plane 
(i.e., $|\tilde{B}_\odot| < 8^\circ$).
For BHBs in GAMA15H ($60^\circ < \tilde{\Lambda}_\odot < 80^\circ$), the Sgr stream is
clearly present from $r=$ 50 kpc to 60 kpc. It is also worth remarking that we identify
the structure labeled as ``feature 3'' in \citet{Sesar2017}. 
It should also be mentioned that the Outer Virgo overdensity
labeled as ``feature 4'' in \citet{Sesar2017}
is absent in our sample, because this structure exists in the region
($\tilde{B}_\odot = -9^\circ$) out of GAMA15H. 
On the other hand, in XMM-LSS ($250^\circ < \tilde{\Lambda}_\odot < 260^\circ$),
there are no stream-like structures because the current BHB sample is located
at larger radii than the Sgr stream (Fig. \ref{fig: stream})

\section{Conclusions}

We have extracted BHB candidates in the early survey data of the HSC-SSP over
$\sim 300$~deg$^2$ based on its $(g,r,i,z)$ photometry, where $z$-band light can be
used as a surface gravity indicator of a star against other contaminants.
Our purpose with selected BHB stars is to trace and map out the Milky Way stellar halo
out to its possible boundary if there is any. 
About 450 BHB candidates have been identified at Galactocentric distances
of 50 to 300~kpc, which corresponds to the $g$-band apparent magnitude of
$g = 19.2 - 23.0$~mag if the absolute $g$-band absolute magnitude $M_g$ of
a BHB is $M_g \simeq 0.7$. Thus, HSC enables to detect BHB stars in the outer part
of the Milky Way halo which no other surveys can reach.

Based on the maximum likelihood method, we have found that the density structure of
the stellar halo at $r = 50-300$~kpc when GAMA15H and XMM-LSS having notable substructures
are excluded is characterized by a single power-law index $\alpha$ of 3.5 and
the axial ratio $q$ of 1.3. This suggests that the stellar halo is slightly prolate
in such an outer halo region.
When we allow a break radius of $r_b$ and different power-law indices inside/outside as
$\alpha_{\rm in}$ and $\alpha_{\rm out}$ for the density profile, we obtain
a steep slope of $\alpha_{\rm out} \simeq 5.3$ outside $r_b \simeq 200$~kpc for the case
without GAMA15H and XMM-LSS.
On the other hand, halo substructures possibly originated from the tidal stream of
the infalling Sgr dwarf dominates the actual halo structure in the outer halo,
making it largely prolate with $q > 1$. 

However, this result is to be assessed using larger BHB sample,
because the outer halo region traced by only a few number of BHBs may be yet
subject to misclassified contaminations such as A stars and background galaxies.
Therefore, the completion of this HSC-SSP survey over $\sim 1,400$~deg$^2$
will be important in assessing the current results with higher statistical
significance and in exploring further structures of the stellar halo in the
Milky Way.

\begin{ack}
We are grateful to the referee for her/his constructive comments and suggestions
that help improve our manuscript substantially.
This work is supported in part by JSPS Grant-in-Aid for Scientific 
Research (B) (No. 25287062) and MEXT Grant-in-Aid for Scientific Research on 
Innovative Areas (No. 15H05889, 16H01086, 17H01101).
N.A. was supported by the Brain Pool Program,
which is funded by the Ministry of Science and ICT
through the National Research Foundation of
Korea (2018H1D3A2000902).

The Hyper Suprime-Cam (HSC) collaboration includes the astronomical
communities of Japan and Taiwan, and Princeton University.  The HSC
instrumentation and software were developed by the National
Astronomical Observatory of Japan (NAOJ), the Kavli Institute for the
Physics and Mathematics of the Universe (Kavli IPMU), the University
of Tokyo, the High Energy Accelerator Research Organization (KEK), the
Academia Sinica Institute for Astronomy and Astrophysics in Taiwan
(ASIAA), and Princeton University.  Funding was contributed by the FIRST 
program from Japanese Cabinet Office, the Ministry of Education, Culture, 
Sports, Science and Technology (MEXT), the Japan Society for the 
Promotion of Science (JSPS),  Japan Science and Technology Agency 
(JST),  the Toray Science  Foundation, NAOJ, Kavli IPMU, KEK, ASIAA,  
and Princeton University.
This paper makes use of software developed for the Large Synoptic Survey Telescope. We thank the
LSST Project for making their code freely available. The Pan-STARRS1 (PS1) Surveys have been made
possible through contributions of the Institute for Astronomy, the University of Hawaii,
the Pan-STARRS Project Office, the Max-Planck Society and its participating institutes, the Max Planck Institute for
Astronomy and the Max Planck Institute for Extraterrestrial Physics, The Johns Hopkins University,
Durham University, the University of Edinburgh, Queen's University Belfast, the Harvard-Smithsonian
Center for Astrophysics, the Las Cumbres Observatory Global Telescope Network Incorporated, the
National Central University of Taiwan, the Space Telescope Science Institute, the National Aeronautics
and Space Administration under Grant No. NNX08AR22G issued through the Planetary Science Division
of the NASA Science Mission Directorate, the National Science Foundation under Grant
No.AST-1238877, the University of Maryland, and Eotvos Lorand University (ELTE).
\end{ack}



\end{document}